\documentstyle[art12,epsf]{article}

\def\QMW{Queen Mary \& Westfield College}

\begin{document}
\begin{titlepage}
\begin{flushright}
INP MSU Preprint-95-1/365\\
QMW-PH-95-6\\
hep-ph/9503401\\
4 May 1995
\end{flushright}

\begin{center}
{\bf\boldmath{
{Single leptoquark production \\
associated with hard photon emission \\ in
$ep$ collisions at high energies}
}}
\end{center}
\bigskip
\begin{center}
{V. Ilyin, A. Pukhov, V. Savrin, A. Semenov, \\
{\small\it Institute of Nuclear Physics of Moscow State University,
	119899 Moscow, Russia}\\
 and \\
 W. von Schlippe \\
{\small\it Queen Mary \& Westfield College, London, England}
 }
\end{center}\bigskip
\vfill
\begin{abstract}
\noindent
In this paper we consider single leptoquark resonance production associated
with the emission of a hard photon.
We have obtained analytical
formul\ae\ for differential and total cross sections
for the cases of scalar and vector leptoquarks.

We have found that in reactions with scalar leptoquarks
there is no photon radiation
in some directions depending on the leptoquark electric charge
({\it the radiative amplitude zero -- the RAZ} effect).

For vector leptoquarks the  exact RAZ is present only in the
case of  Yang-Mills coupling.
We propose to use the RAZ effect to determine the types of leptoquarks.
We conclude also that this effect
opens a possibility to measure the leptoquark anomalous magnetic moment.
\end{abstract}

\vfill
\end{titlepage}

\section{Introduction.}

During the last ten years many authors have considered the existence of
leptoquarks as a possible extension of the Standard Model (SM)
(see, for example, \cite{LQ-models} and references therein). Such schemes are
attractive because of the similarity  of quarks and
leptons with respect to the structure of electroweak interaction.
Some fundamental problems of quantum field theory can be
resolved with the help of such new bosons.
A recent example is the attempt to explain the
unification of the three gauge couplings at the
grand unified theory scale \cite{GUT-SM} by
introducing a scalar leptoquark isodoublet  into the model.
So it appears to be
quite interesting to admit the possibility for
the quark to convert into a
lepton and back due to the emission of a new boson --- the {\it leptoquark}.
These hypothetical bosons are colour particles with lepton and baryon
numbers. In this paper we refer to the leptoquarks as LQ.

Various general restrictions on possible sets of these new particles and
their properties
follow from low-energy experiments (proton stability,
rare decays, quark-lepton universality etc.), see \cite{BW86,BRW,KM-LQ}.
Among them are
(i) baryon and lepton number conservation;
(ii) chirality of fermion-LQ couplings ;
(iii) some ``fermion family'' restrictions.

The principal mode of LQ production
at electron-proton colliders is electron-quark fusion
which gives rise to resonance peaks in the cross-section of $ep$ collisions.
The experimental investigations of such collisions at high energies allow
one to measure LQ masses directly
through the resonance peak position
and to establish bounds in the
(LQ mass, fermion-LQ coupling ($\lambda$)) plane.
Direct searches for new particles by their s-channel resonances in
the H1 and ZEUS experiments at HERA have been presented in Ref. \cite{HERA}.
The characteristic bounds for the LQ masses at
$\lambda=0.3$ lie in the interval $140$ to $235\,\mbox{GeV}$
depending on the LQ type.
 These bounds were derived from the analyses of
data samples of $\approx 425 \,\mbox{nb}^{-1}$ accumulated
by each of the two HERA experiments.
With larger accumulated luminosity the whole kinematical
range, up to $296\,\mbox{GeV}$, will be probed for the existence of
LQ resonances.
The next step in $ep$-collision experiments
could be the LEP+LHC project with
$\sqrt{s}=1740\,\mbox{GeV}$ and an annual
luminosity about 1 fb$^{-1}$.

However a new important problem will arise after the observation of the LQ
signal, namely the problem of identifying  the leptoquark: it will be
necessary to measure not only the LQ masses but also
the fermion-LQ couplings and
quantum numbers, such as electric charge, hypercharge etc.

In this paper we show a possibility to carry out the
LQ identification using data from $ep$ experiments.
For this purpose we propose to investigate single LQ
production associated with the emission of a hard photon.
We have found that in this reaction there is no photon radiation
in some directions depending on the LQ type. Following Ref.
\cite{raz-ref} we will call this effect
{\it radiative amplitude zero (RAZ)}.
The value of the RAZ angle depends essentially on the LQ electric charge.
As a result the RAZ test gives us a possibility to determine the LQ
type.

 From a phenomenological point of view it is reasonable to consider
the most general Lagrangian for fermion-LQ vertices
satisfying the general
restrictions listed above, with
dimensionless couplings, and satisfying
$SU(3)_c\times SU(2)_L\times U(1)_R$ gauge invariance. In this
case a general phenomenological analysis can be
made without reference to
any concrete model. Such a Lagrangian was proposed in
Ref. \cite{BRW} for the first generation of fermions.

The interaction of scalar LQ with electromagnetic field
can be described by a Lagrangian
which is fixed by its electric charge:
$$ {\cal L}^S_{\gamma} = (D^\mu \Phi)^+(D_\mu \Phi)-M^2\Phi^+\Phi. $$
Here
$D_\mu=\partial_\mu-ieQA_\mu$,
$A_\mu$ is the electromagnetic field,
and we denote a scalar LQ by $\Phi$.

The Lagrangian for vector LQ interaction
with electromagnetic field can be written in the following form:
\begin{eqnarray}
{\cal L}^V_\gamma =
ieQ[A_\mu\Phi^{+\mu\nu}\Phi_\nu &-&
 A_\mu\Phi_\nu^+\Phi^{\mu\nu}
 +(1-\kappa) A_{\mu\nu}\Phi^{+\nu}\Phi^\mu] \nonumber \\
 &-&(eQ)^2[A^2(\Phi^+_\mu\Phi^\mu)-(A^\mu\Phi^+_\mu)(A^\nu\Phi_\nu)].
                                              \label{eq:vLQ-lagr}
\end{eqnarray}
Here $\Phi_{\mu\nu}=\partial_\nu\Phi_\mu-\partial_\mu\Phi_\nu$,
$A_{\mu\nu}=\partial_\nu A_\mu-\partial_\mu A_\nu$,
and the vector LQ is denoted by $\Phi_\mu$.

The LQ quantum numbers are collected in Table \ref{tab:LQ-Zoo} and
correspond to Ref. [4].
Here the third components of isospin
are denoted by $T_3$ and the LQ electric charges by $Q$.
In the following text we use notation, like $S^{-1}_3$, corresponding to
$(LQ)^{T_3}_{2T+1}$, where $T$ is the $SU(2)$ isospin.

Since one of the general low-energy restrictions is the
chirality of the fermion-LQ coupling,
we consider in our calculations only either left-handed couplings
 or right-handed ones
 using the same notation $\lambda$ in both cases.

In Lagrangian (\ref{eq:vLQ-lagr})
there is a free dimensionless parameter $\kappa$,
the {\it anomalous magnetic moment} of the LQ. The value
$\kappa=0$ corresponds to the Yang-Mills structure
and $\kappa=1$ corresponds to {\it minimal coupling}.

We have carried out our analytical calculations using the CompHEP
package \cite{CompHEP}.

 \section{Single leptoquark production}

To study the LQ interaction  with electromagnetic
field we propose the following inclusive reaction:
\begin{equation}
 e^\pm +  p\rightarrow \gamma +LQ+X
                                               \label{eq:gammLQ-i}
\end{equation}
with a hard photon in the final state.

An obvious  starting point for the analysis
of reaction (\ref{eq:gammLQ-i})
is the assumption that a sharp peak is observed in the reaction
\begin{equation}
 e^\pm + p \rightarrow \ell+q+X .
                                                \label{eq:resLQ-i}
\end{equation}
This peak is a signal of the LQ resonant electron (positron) - quark fusion
\begin{equation}
 e^\pm + q\rightarrow LQ.
                                                \label{eq:eq-LQ-eq}
\end{equation}

In such a situation the mass of the new particle can be
determined directly from the position of the sharp peak in the
$x$ distribution of the struck quark \cite{BRW}.

However in the literature there is no detailed investigation of a problem of
determining other quantum numbers of the observed LQ.
In this paper we propose a solution of this problem based on data
 from ep-collisions. Certainly, data from the collisions of other types can
give necessary information (see, for example, \cite{Blum-Boos} and
references therein).

First, we note that one can determine the lepton chirality of
the observed LQ by using polarized electron (positron)
beams.

The spin of the new particle can be easily determined from
reaction (\ref{eq:resLQ-i}): for scalar LQ
the distribution in the decay polar angle $\theta^*$ is isotropic
in the LQ rest frame, whereas
for vector LQ it has a characteristic
$(1+\cos \theta^*)^2$ dependence.

However it will be practically impossible to determine
which flavour of quark or antiquark has produced this leptoquark.
Indeed, the total resonance production cross section
(\ref{eq:eq-LQ-eq}) is given by (cf. Ref. \cite{BRW})
$$
\sigma_{tot} (e^\pm p\rightarrow LQ + X)=
\frac{\pi}{4s}\cdot\lambda^2\cdot f_q (\frac{M^2}{s})\cdot(J+1),
$$
where $J$ is the LQ spin ($J = 0,\, 1$),
and $f_q (x)$ is the quark (antiquark) distribution function.
We see that the value of the total cross section allows one to
measure the product $\lambda^2 f_q(x)$ rather than the parton structure
function at $x={M^2}/{s}$ and $\lambda$ separately. So in this way we cannot
conclude what proton constituent was involved in the LQ production.

Furthermore, one could think of measuring the coupling constant
$\lambda$ directly from the shape of the observed resonance peak using the
formul\ae\ for the partial widths \cite{BRW} in the scalar and vector cases
correspondingly:
$$
\Gamma_{LQ}^S=\frac{\lambda^2}{16\pi}\cdot M,
\qquad
\Gamma_{LQ}^V=\frac{\lambda^2}{24\pi}\cdot M.
$$
Unfortunately, even for the electroweak value of the coupling constant,
$\lambda\sim 0.3$, the partial
width is too small, $\sim 0.002 M$, to be resolved experimentally.

Another possibility to consider is the direct measurement of
quark-jet charges event-by-event to assist in the identification
of the LQ type. However, whereas existing methods work reasonably
well for heavy quarks, they do require very large statistics in
the case of light quarks (see, for example, \cite{j-charge}).
Also, the known methods appear to work only for integrated
characteristics, such as for instance asymmetry and partial widths.

We conclude that the analysis of resonant LQ production in $ep$ collisions
cannot help us in  identifying  the type of the observed new particle
and, as a consequence, in measuring the fermion-LQ coupling constant
$\lambda$. In the following we show the potential of reaction
(\ref{eq:gammLQ-i}) in the identification of the observed new particle.
We will show that it is essential to exploit the RAZ
effect for this purpose.

\section{Parton cross sections. Analytical results \label{analform}}

In reaction (\ref{eq:gammLQ-i}) there are the following
contributing exclusive subprocesses
\begin{equation}
e^\pm + q\rightarrow \gamma +LQ,
\label{eq:eq-gammLQ}
\end{equation}
where q is a constituent quark of the proton. We consider the case
when the LQ interacts only with fermions of the first generation, i.e.
$q=(u,\bar u, d, \bar d)$.

The amplitude for subprocess (\ref{eq:eq-gammLQ}) is represented by
three Feynman diagrams, see Fig. 1.

The contribution of a separate constituent quark in the
integrated cross section of reaction (\ref{eq:gammLQ-i})
is expressed as the convolution of the subprocess cross section
with the corresponding parton distribution function:
\begin{equation}
\sigma(s)\;=\; \int^1_{x_{min}} dx\; q(x,Q^2)\, \int^1_{-1}
    d\cos{\vartheta_\gamma}\cdot
   \frac{d\hat\sigma(\hat s,\cos{\vartheta_\gamma})}{d\cos{\vartheta_\gamma}}
   \cdot\Theta_{cuts}(E_{\gamma},\vartheta_\gamma).
                                               \label{eq:i-sigma-gammLQ}
\end{equation}
Here $\hat\sigma$ is the cross section of the
corresponding subprocess (\ref{eq:eq-gammLQ}), $\hat s=xs$,
the quark distribution function is denoted by $q(x,Q^2)$
and the four-momentum transfer scale is taken to be $Q^2 = \hat s$.
We denote the photon emission angle by $\vartheta_\gamma$. The direction
$\vartheta_\gamma=0$ is along the proton beam.
The function $\Theta_{cuts}(E_{\gamma},\vartheta_\gamma)$ introduces the
necessary kinematical cuts. The process under discussion is
infra-red divergent, therefore
we have to introduce the cut $E_\gamma > E^0_\gamma >0$.

We have calculated analytically the unpolarized squared matrix elements and
cross sections for subprocesses (\ref{eq:eq-gammLQ}).
The electron and quark ($u$ and $d$) masses are taken equal to zero in
these calculations.

The squared amplitude for subprocesses (\ref{eq:eq-gammLQ}) in the
case of scalar leptoquarks can be written in the following
form\footnote{Results in the case of $S_3^{\pm 1}$  leptoquarks
are greater by a factor of two.}:
\begin{equation}
|A|^2_S\; =\;\frac{e^2\lambda^2}{2}\cdot
      \frac{\xi^2+1}{(\xi-1)^2}\cdot
      \frac{(Q_e \upsilon-Q_q \tau)^2}{\upsilon \tau}.
                                                      \label{eq:sLQ-A2}
\end{equation}
Here q denotes the incoming quark or antiquark
($u$, $d$, $\bar u$, or $\bar d$),
$e$ is the elementary charge, $Q_q$ and $Q_e = \mp 1$ are
the quark, electron and positron charges respectively, in units of $e$.
We also use normalized dimensionless Mandelstam variables
$$
\xi\equiv \frac{\hat s}{M^2}\;=\;\frac{(p_e + p_q)^2}{M^2},
 \;\;\;\;\;\;
\tau\equiv \frac{t}{M^2}\;=\;\frac{(p_{\gamma} - p_e)^2}{M^2}, \;\;\;\;\;\;
\upsilon\equiv \frac{u}{M^2}\;=\;\frac{(p_{\small LQ} - p_e)^2}{M^2}.
$$

The total cross sections of the subprocesses (with scalar LQ)
are given by

$$
  \hat\sigma^S(\xi,\delta_1,\delta_2)\;=\;
  \frac{\alpha\lambda^2}{8 M^2}
  \cdot \frac{1}{\xi^2} \cdot
  \frac{\xi^2 +1}{(\xi -1)^2} \cdot
  \left[ L(\xi -1)  - Q^2 (\xi -1 - \delta_1 - \delta_2) \right].
$$
Here $\alpha=e^2/(4\pi)$
is the fine structure constant and $Q=Q_e+Q_q$ is the LQ electric charge.
In this formula (also in the corresponding formula for vector LQ)
$$
L\;\equiv\; \log{\frac{\xi -1-\delta_2}{\delta_1}} +
       Q^2_q \log{\frac{\xi -1-\delta_1}{\delta_2}}
$$
and we have applied the cuts on the momentum transfer $\tau$:
\begin{equation}
 -\xi+1+\delta_2\;<\;\tau\;<\;-\delta_1,\qquad \delta_{1,2}>0,
                                                 \label{eq:d1d2cut}
\end{equation}

The squared matrix element for vector LQ has the form\footnote{Results
in the case of  $U_3^{\pm 1}$ leptoquarks
are greater by a factor of two.}
\begin{equation}
|A|^2_{V}=\frac{e^2\lambda^2}{(\xi -1)^2}\cdot
\left[(Q_q \tau-Q_e \upsilon)^2 K_0\,+\,
Q (Q_q \tau-Q_e \upsilon) K_1 \kappa\;+\;
Q^2 K_2\kappa^2\right],
                                              \label{eq:vLQ-A2}
\end{equation}

$$
  K_0\equiv \frac{{\xi}^2+1-2\tau\upsilon}{\tau\upsilon},\qquad
  K_1\equiv \tau-\upsilon,\qquad
  K_2\equiv \frac{\tau\upsilon}{2}\; +\;
        \frac{\xi}{8}\cdot (\tau^2+\upsilon^2).
$$

The total cross sections of the subprocesses with vector LQ and
with the cuts on momentum transfer (\ref{eq:d1d2cut}) applied can be
represented in the following form

\begin{equation}
 \hat\sigma^V(\xi,\delta_1,\delta_2) =
   \frac{\alpha\lambda^2}{4 M^2 \xi^2 (\xi-1)^2}
 \left\{ (\xi^2+1)(\xi-1) L
   + \sum^3_{i=0} (A_i+B_i\kappa+C_i\kappa^2) \xi^i \right\}.
\end{equation}

Here

$$
A_0= 2 - 2 Q_e Q + \frac{5}{3} Q^2
      + \delta_1 (Q^2 + 2) + \delta_2 (Q^2 + 2 Q_q^2)
      + 2 (\delta_1^2 Q_e+\delta_2^2 Q_q) Q
      + \frac{2}{3} (\delta_1^3+\delta_2^3) Q^2; $$
$$
A_1= - 3 (Q_q^2+1) - 4 (\delta_1 +  \delta_2 Q_q^2 )
     - 2 (\delta_1^2 Q_e+\delta_2^2 Q_q) Q; $$
$$
A_2= 3 (Q_q^2+1)+\delta_1 (Q^2+2)+\delta_2 (Q^2+2 Q_q^2); \qquad
A_3 =-2 + 2 Q_e Q - \frac{5}{3} Q^2;
$$


$$
B_0=-\frac{Q}{6}
\left[ Q + 6 (\delta_1 Q_e+\delta_2 Q_q)
+ 3 (\delta_1^2+\delta_2^2) Q + 6 (\delta_1^2 Q_e+\delta_2^2 Q_q)
+4 (\delta_1^3+\delta_2^3) Q \right]; $$
$$
B_1=\frac{Q}{2} \left[ Q + 4 (\delta_1 Q_e + \delta_2 Q_q)
+(\delta_1^2+\delta_2^2) Q + 2 (\delta_2^2 Q_q + \delta_1^2 Q_e)\right];
$$
$$
B_2=-\frac{Q}{2} \left[Q+2(\delta_1 Q_e+\delta_2 Q_q)\right]; \qquad
B_3= \frac{Q^2}{6};
$$


$$
C_0= -\frac{Q^2}{12}
\left[1-3 (\delta_1^2+\delta_2^2)-2 (\delta_1^3+\delta_2^3)\right]; $$
$$
C_1=\frac{Q^2}{24}
\left[4-3(\delta_1+\delta_2)
       -9 (\delta_1^2+\delta_2^2)-2 (\delta_1^3+\delta_2^3)\right]; $$
$$
C_2= \frac{Q^2}{8}
\left[2 (\delta_1 + \delta_2)+ \delta_1^2 + \delta_2^2\right]; \qquad
C_3= -\frac{Q^2}{24} \left[4 +3 (\delta_1+\delta_2)\right];
$$

Consider these formul\ae\ in connection with the RAZ effect.
The factor $(Q_e \upsilon-Q_q \tau)^2$ in Eqs.
(\ref{eq:sLQ-A2},\ref{eq:vLQ-A2}) gives
the RAZ, i.e. the absence of photon emission
in some direction. Note that for vector leptoquarks the exact RAZ is present
only in the Yang-Mills case ($\kappa=0$)  due to the $\kappa^2$ term.

The value of the corresponding polar angle is determined by
the relation $Q_e \upsilon=Q_q \tau$.
We have the
formula for the RAZ angle in the CMS of the pair ($e^\pm_{in}$, $q_{in}$)
$$\cos\vartheta^*_{RAZ}=\frac{Q_e-Q_q}{Q}. $$

Due to different values of LQ charges the RAZ effect
exists only for some of the leptoquark types collected in
Table \ref{tab:RAZ-angle}.
As a result, we have a possibility to identify
the type of the discovered leptoquark by studying
the distribution in the photon emission angle.

In  Fig. \ref{fig:eq-RAZ} we show the distributions in the photon
emission angle in the CMS of the ($e^\pm_{in}$, $q_{in}$) pair.

In the laboratory frame the RAZ angle is changed by the corresponding
Lorentz boost. This boost depends on $\hat s = xs$, so the RAZ is shifted and
smoothed due to the distribution in $x$. However this
distribution has a rather sharp peak for a small enough cut over the photon
energy.
As a result, for the rough estimates we can use $\hat s= M^2$,
and in order to estimate the RAZ angle in the laboratory frame we can use
the following formul\ae\
$$
   \cos{\vartheta_{RAZ}}\;=\;
   \frac{\cos{\vartheta^*_{RAZ}} + \hat v}
   {1+\hat v \cdot\cos{\vartheta^*_{RAZ}}},
   \qquad \hat v\;=\; \frac{M^2 - 4 {\cal E}^2_e}{M^2 + 4 {\cal E}^2_e}.
$$
Here the electron beam energy is denoted by ${\cal E}_e$.

Note that at HERA the RAZ is shifted significantly whereas at
LEP+LHC the case of
$M=200\,\mbox{GeV}$ practically corresponds to the CMS.
We give some RAZ angles for different LQ masses
in Table \ref{tab:RAZ-angle}.

A final remark is that cross sections  with the RAZ
effect in the case of vector LQ are sensitive
to the value of the LQ anomalous magnetic moment,
the exact RAZ being present only for the Yang-Mills type of
photon-leptoquark coupling.  This effect opens therefore a
possibility to measure the LQ anomalous magnetic moment.

\section*{Conclusions}

In this paper we have analysed the formulas obtained with the help of the
CompHEP package for the process of single leptoquark production associated
with hard photon emission which reveal the absence of radiation at some
photon emission angles. We propose to use this so-called radiative amplitude
zero (RAZ) effect as a tool for identifying the leptoquarks that could
manifest themselves as resonance peaks in $ep$ collisions. We have shown that
for vector leptoquarks the exact RAZ effect is present only for the
Yang-Mills couplings. This opens a possibility to probe the leptoquark
anomalous magnetic moment.

Here we have concentrated on the analytical results for
subprocesses with two particles in the final state.
However preliminary numerical
calculations show that the same conclusions can be drawn for the
more realistic
case taking into account parton distributions and three body final states.
The results of accurate numerical calculations will be presented in future
publications.

\section*{Acknowlegements}

This work was partly supported by The Royal Society of London as a
joint project between Queen Mary \& Westfield College (London) and the
Institute of Nuclear Physics of Moscow State University, and
by INTAS (project 93-1180).
V.I. and V.S. wish to thank \QMW\ for its hospitality and
the possibility to work during their visit in London.
WvS wishes to thank the Nuclear Physics Institute of
Moscow State University for its hospitality.
The work of V.I., A.P. and A.S. was partly supported also by the Russian
Foundation for Fundamental Research (RFFI grant 93-02-14428)
and by Grant M9B000 from the International Science Foundation.
We thank E. Boos, J. Fujimoto, I. Ginzburg and Y. Shimizu for useful
discussions.

\eject

\section*{Tables}

\begin{table}[h]
{\small
\begin{center}
\begin{tabular}{|cccccc|rr|}
\hline
\multicolumn{1}{|c}{\small \it LQ} &
\multicolumn{1}{|c|}{\small\it Lepton} &
\multicolumn{1}{c|}{\small \it Spin}&
\multicolumn{1}{c}{\small \it F} &
\multicolumn{1}{|c|}{\small \it $SU(3)_c$} &
\multicolumn{1}{c|}{\small \it Y=} &
\multicolumn{1}{c}{\small \it $ T_3 $} &
\multicolumn{1}{|c|}{\small \it $Q$ } \\
\multicolumn{1}{|c}{\small \it name} &
\multicolumn{1}{|c|}{\small \it chirality} &
\multicolumn{1}{|c}{ } &
\multicolumn{1}{|c}{ } &
\multicolumn{1}{|c|}{ } &
\multicolumn{1}{|c|}{\small \it $2(Q-T_3)$} &
\multicolumn{1}{c}{ } &
\multicolumn{1}{|c|}{ } \\ \hline
\raisebox{0ex}[3ex][2ex]{$S_1$}
      &L,R&0&2& 3      &$-\frac{2}{3}$ & 0            &$-\frac{1}{3}$\\
\hline
\raisebox{0ex}[3ex][2ex]{$\tilde S_1$}
&R  &0&2& 3      &$-\frac{8}{3}$ & 0            &$-\frac{4}{3}$\\
\hline
              & & & & & &-1&\raisebox{0ex}[3ex][2ex]{$-\frac{4}{3}$}\\
                                                         \cline{7-8}
\raisebox{0ex}[3ex][2ex]{$S_3$} &L&0&2&3&$-\frac{2}{3}$& 0&$-\frac{1}{3}$\\
                                                         \cline{7-8}
              & & & & & & 1&\raisebox{0ex}[3ex][2ex]{$\frac{2}{3}$} \\
\hline
\raisebox{0ex}[3ex][2ex]{$R_2$}&L,R&0&0&$\bar 3$
                         &$-\frac{7}{3}$&$-\frac{1}{2}$&$-\frac{5}{3}$\\
                                                         \cline{7-8}
  & & & & & &$\frac{1}{2}$&\raisebox{0ex}[3ex][2ex]{$-\frac{2}{3}$}\\
\hline
\raisebox{0ex}[3ex][2ex]{$\tilde R_2$}&L  &0&0&$\bar 3$&$-\frac{1}{3}$
 &$-\frac{1}{2}$&$-\frac{2}{3}$\\
                                                         \cline{7-8}
  & & & & & &$\frac{1}{2}$&\raisebox{0ex}[3ex][2ex]{$\frac{1}{3}$}\\
\hline
\raisebox{0ex}[3ex][2ex]{$U_1$}       &L,R&1&0&$\bar 3$&$-\frac{4}{3}$
 & 0            &$-\frac{2}{3}$\\
\hline
\raisebox{0ex}[3ex][2ex]{$\tilde U_1$}&R  &1&0&$\bar 3$&$-\frac{10}{3}$
& 0            &$-\frac{5}{3}$\\
\hline
                  & & & & & &-1&\raisebox{0ex}[3ex][2ex]{$-\frac{5}{3}$}\\
                                                     \cline{7-8}
\raisebox{0ex}[3ex][2ex]{$U_3$}&L&1&0&$\bar 3$
&$-\frac{4}{3}$&0&$-\frac{2}{3}$\\
                                                     \cline{7-8}
                  & & & & & & 1&\raisebox{0ex}[3ex][2ex]{$\frac{1}{3}$}\\
\hline
\raisebox{0ex}[3ex][2ex]{$V_2$}
 &L,R&1&2& 3      &$-\frac{5}{3}$ &$-\frac{1}{2}$&$-\frac{4}{3}$\\
\cline{7-8}
  & & & & & &$\frac{1}{2}$&\raisebox{0ex}[3ex][2ex]{$-\frac{1}{3}$}\\
\hline
\raisebox{0ex}[3ex][2ex]{$\tilde V_2$}&L  &1&2
& 3      &$\frac{1}{3}$  &$-\frac{1}{2}$&$-\frac{1}{3}$\\
\cline{7-8}
  & & & & & &$\frac{1}{2}$&\raisebox{0ex}[3ex][2ex]{$\frac{2}{3}$}\\
\hline
\end{tabular}
\end{center}
}
\caption{Leptoquark quantum numbers.}
\label{tab:LQ-Zoo}
\end{table}
\clearpage

\begin{table}[h]
\large
\begin{center}
\begin{tabular}{|l|c|c|c|}
\hline
& M & \multicolumn{2}{c|}{$\vartheta_{RAZ}$ [{\small \it deg}]} \\
\cline{3-4}
&[{\small\it GeV}]
&$\vartheta^*_{RAZ}=60^\circ$&$\vartheta^*_{RAZ}=78.5^\circ$ \\
\cline{3-4}
&& \raisebox{0ex}[3ex][2ex]{$S_3^{-1}$}, $V_2^{-{1\over 2}}$ (L)
& $R_2^{-{1\over 2}}$, $U_3^{-1}$ (L)   \\
      &         &   $\tilde S_1$, $V_2^{-{1\over 2}}$ (R)
& $R_2^{-{1\over 2}}$, $\tilde U_1$ (R)\\ \hline
HERA  & 150     &   26       &  36 \\
      & 200     &   20       &  27 \\
      & 250     &   16       &  22 \\ \hline
LEP+LHC & 200   &   60        &   78.5 \\
      &  300    &   42      &  57  \\
      &  500    &   26      &  36  \\
      & 1000    &   13      &  18  \\
      & 1500    &   9       &  13  \\ \hline
\end{tabular}
\end{center}
\caption{RAZ angles in dependence on LQ mass.}
\label{tab:RAZ-angle}
\end{table}
\clearpage

\eject

\section*{Figures}

\def\diagA{
\begin{picture}(170,220)(0,0)
\thinlines
\put(72.9,111.4){\vector(1,0){0}}
\put(34.3,111.4){\makebox(0,0)[r]{$e^-$}}
\put(37.1,111.4){\line(1,0){71.4}}
\put(147.1,111.4){\makebox(0,0)[l]{$\gamma$}}
\multiput(108.1,111.4)(4.7,0.0){8}{\rule[-0.5pt]{1.0pt}{1.0pt}}
\multiput(108.6,111.4)(4.7,0.0){8}{\rule[-0.5pt]{1.0pt}{1.0pt}}
\multiput(109.1,111.4)(4.7,0.0){8}{\rule[-0.5pt]{1.0pt}{1.0pt}}
\multiput(109.6,111.4)(4.7,0.0){8}{\rule[-0.5pt]{1.0pt}{1.0pt}}
\multiput(110.1,111.4)(4.7,0.0){8}{\rule[-0.5pt]{1.0pt}{1.0pt}}
\put(108.6,75.7){\vector(0,-1){0}}
\put(102.9,75.7){\makebox(0,0)[r]{$e^-$}}
\put(108.6,111.4){\line(0,-1){71.4}}
\put(72.9,40.0){\vector(1,0){0}}
\put(34.3,40.0){\makebox(0,0)[r]{$q$}}
\put(37.1,40.0){\line(1,0){71.4}}
\put(126.4,40.0){\vector(1,0){0}}
\put(147.1,40.0){\makebox(0,0)[l]{$LQ$}}
\multiput(108.1,40.0)(3.2,0.0){12}{\rule[-0.5pt]{1.0pt}{1.0pt}}
\end{picture}
}

\def\diagB{
\begin{picture}(170,220)(0,0)
\thinlines
\put(55.0,93.6){\vector(1,-1){0}}
\put(34.3,111.4){\makebox(0,0)[r]{$e^-$}}
\put(37.1,111.4){\line(1,-1){35.7}}
\put(55.0,57.9){\vector(1,1){0}}
\put(34.3,40.0){\makebox(0,0)[r]{$q$}}
\put(37.1,40.0){\line(1,1){35.7}}
\put(90.7,75.7){\vector(1,0){0}}
\put(90.0,84.3){\makebox(0,0){\scriptsize $LQ$}}
\multiput(72.4,75.7)(3.2,0.0){12}{\rule[-0.5pt]{1.0pt}{1.0pt}}
\put(126.4,93.6){\vector(1,1){0}}
\put(147.1,111.4){\makebox(0,0)[l]{$LQ$}}
\multiput(108.1,75.7)(3.2,3.2){12}{\rule[-0.5pt]{1.0pt}{1.0pt}}
\put(147.1,40.0){\makebox(0,0)[l]{$\gamma$}}
\multiput(108.1,75.7)(4.7,-4.7){8}{\rule[-0.5pt]{1.0pt}{1.0pt}}
\multiput(108.6,75.2)(4.7,-4.7){8}{\rule[-0.5pt]{1.0pt}{1.0pt}}
\multiput(109.1,74.7)(4.7,-4.7){8}{\rule[-0.5pt]{1.0pt}{1.0pt}}
\multiput(109.6,74.2)(4.7,-4.7){8}{\rule[-0.5pt]{1.0pt}{1.0pt}}
\multiput(110.1,73.7)(4.7,-4.7){8}{\rule[-0.5pt]{1.0pt}{1.0pt}}
\end{picture}
}

\def\diagC{
\begin{picture}(170,220)(0,0)
\thinlines
\put(72.9,111.4){\vector(1,0){0}}
\put(34.3,111.4){\makebox(0,0)[r]{$e^-$}}
\put(37.1,111.4){\line(1,0){71.4}}
\put(126.4,111.4){\vector(1,0){0}}
\put(147.1,111.4){\makebox(0,0)[l]{$LQ$}}
\multiput(108.1,111.4)(3.2,0.0){12}{\rule[-0.5pt]{1.0pt}{1.0pt}}
\put(108.6,75.7){\vector(0,1){0}}
\put(102.9,75.7){\makebox(0,0)[r]{$q$}}
\put(108.6,111.4){\line(0,-1){71.4}}
\put(72.9,40.0){\vector(1,0){0}}
\put(34.3,40.0){\makebox(0,0)[r]{$q$}}
\put(37.1,40.0){\line(1,0){71.4}}
\put(147.1,40.0){\makebox(0,0)[l]{$\gamma$}}
\multiput(108.1,40.0)(4.7,0.0){8}{\rule[-0.5pt]{1.0pt}{1.0pt}}
\multiput(108.6,40.0)(4.7,0.0){8}{\rule[-0.5pt]{1.0pt}{1.0pt}}
\multiput(109.1,40.0)(4.7,0.0){8}{\rule[-0.5pt]{1.0pt}{1.0pt}}
\multiput(109.6,40.0)(4.7,0.0){8}{\rule[-0.5pt]{1.0pt}{1.0pt}}
\multiput(110.1,40.0)(4.7,0.0){8}{\rule[-0.5pt]{1.0pt}{1.0pt}}
\end{picture}
}

\begin{figure}[h]
\unitlength=0.7pt
\begin{center}
\diagB\ \diagA\ \diagC
\end{center}
\caption{Feynman diagrams for subprocesses $e^- + q\rightarrow\gamma +LQ$}
\label{fig:gammLQ-diagr}
\end{figure}

\unitlength=1cm

\begin{figure}[h]
\begin{picture}(16,5)
\put(6,0){\epsfxsize=7.5cm \leavevmode \epsfbox{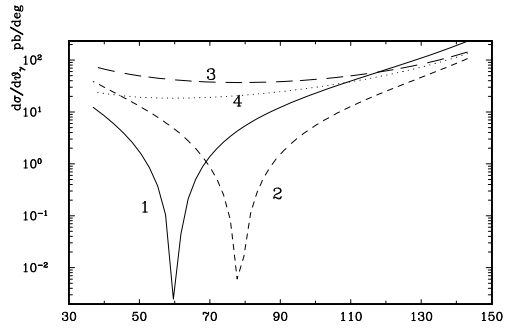} }
\end{picture}
\caption{Angular distributions in the CMS for scalar LQ at
 $M_{LQ}=300\,\mbox{GeV}$,
$\hat s=(304\,\mbox{GeV})^2$ and $\lambda=0.3$.
1) $e^- + d\to\gamma+S^{-1}_3$; 2) $e^+ + u\to\gamma+R_2$;
3) $e^- + u\to\gamma+S_1(S^0_3)$; 4) $e^+ + d\to\gamma+\tilde R_2$}
\label{fig:eq-RAZ}
\end{figure}

\end{document}